\documentclass[usenatbib]{mn2e}
\usepackage{graphicx}

\title[Site Testing]{Detection of a 14-days atmospheric perturbation peak at Paranal associated with lunar cycles
}

\author[S. Cavazzani et al.]{S. Cavazzani$^{1,2}$\thanks{E-mail:stefano.cavazzani@unipd.it}, S. Ortolani$^{1,2}$, N. Scafetta$^{3}$, V. Zitelli $^{4}$, G. Carraro$^{1}$
 \\
$^{1}$Department of Physics and Astronomy, University of Padova, Vicolo
dell'Osservatorio 3, 35122, Padova, Italy\\
$^{2}$INAF-Osservatorio Astronomico di Padova, Vicolo dell'Osservatorio 5, 35122, Padova, Italy\\
$^{3}$Dipartimento di Scienze della Terra, dell'Ambiente e delle Risorse, University of Naples, Via Cinthia 21, 80126, Naples, Italy \\
$^{4}$INAF-OAS Osservatorio di Astrofisica e Scienza dello Spazio di Bologna, Via Gobetti 93/3, 40129, Bologna, Italy
}

\begin{document}

\date{Accepted 0000 September 00.  Received 0000 September 00; in original form 0000 May 00.}

\pagerange{\pageref{firstpage}--\pageref{lastpage}} \pubyear{2009}

\maketitle

\label{firstpage}

\begin{abstract}

In this paper we investigate the correlation between the atmospheric perturbations at Paranal Observatory and the Chilean coast tides, which are mostly modulated by the 14-day syzygy solar-lunar tidal cycle.
To this aim, we downloaded 15 years (2003-2017) of cloud coverage data from the AQUA satellite, in a matrix that includes also Armazones, the site of the European Extremely Large Telescope.
By applying the Fast Fourier Transform to these data we detected a periodicity peak of about 14 days.
We studied the tide cycle at Chanaral De Las Animas, on the ocean coast, for the year 2017, and we correlated it with the atmospheric perturbations at Paranal and the lunar phases.
We found a significant correlation ($96\%$) between the phenomena of short duration and intensity (1-3 days) and the tidal cycle at Chanaral. 
We then show that an atmospheric perturbation occurs at Paranal in concomitance with the low tide, which anticipates the full (or the new) moon by 3-4 days. This result allows to improve current weather forecasting models for astronomical observatories by introducing a lunar variable.

\end{abstract}

\begin{keywords}
 atmospheric effects -- optical turbulence -- tidal atmospheric influence.
\end{keywords}

\section{Introduction}

\begin{figure}
  \centering
  \includegraphics[width=8cm]{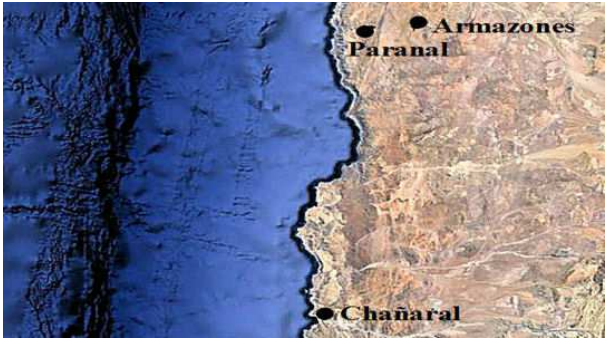}
  \caption{Location of the analyzed sites, the matrix $1^{\circ}\times 1^{\circ}$ used for the analysis of atmospheric perturbations and pressure contains both sites: Paranal (LAT. $-24^{\circ}37'$, LONG. $-70^{\circ}24'$, Altitude 2635 m) and Armazones (LAT. $-24^{\circ}35'$, LONG. $-70^{\circ}11'$, Altitude 3064 m). Chanaral De Las Animas (LAT. $-26^{\circ}20'$, LONG. $-70^{\circ}36'$, Altitude 198 m) was used for the oceanic tides analysis.}
\label{i00}
\end{figure}

Reliable predictions of the observational conditions at astronomical sites, especially the degree of cloud coverage, are a crucial ingredient for modern astronomical observations. 
The widespread use of adaptive optics in particular is very sensitive to the cloud coverage conditions (as well as to strong wind and seeing conditions) for a proper operation. 
Both the optimisation of the resources and the observational scheduling program require the statistics of clear/mixed or covered time and its yearly, monthly or shorter time distribution. In addition, recent requests for optical communications for astronomical applications (with satellites or between distant telescopes) make the cloud coverage predictions more relevant.  
Generally speaking, there are two approaches to this issue. One consists in the short time forecast which predicts the cloud coverage level up to a few days, while the other is based on the statistical analysis of longer term trends. Example of short time forecasts of observing conditions (including wind and turbulence) are in Masciadri et al. (2002), Sarazin (2005), Giordano et al. (2013), Cavazzani et al. (\cite{cava15}), and Osborn et al. (\cite{Osborn}). On the other hand, the availability of longer term data from satellites  permits a statistical approach and the search for periodic trends. 
In this paper we show an analysis of long term forecast making use of the Cerro Paranal Observatory environmental data for a detailed temporal analysis of atmospheric perturbations, including  cloud cover,  strong wind, high humidity, and bad seeing conditions. The model proposed in this paper correlates moon phases, tides and atmospheric perturbations to forecast observing required conditions on long and short terms. In particular, we seek empirical confirmations of a possible lunisolar tidal modulation of meteorological parameters,  which has been the subject of lively debate (Crawford, 1982, Lakshmi et al., 1998, and Hagan et al., 2003).
We analysed a  15 year database of cloud cover at Paranal (2003-2017) from the MODIS (Moderate Resolution Imaging Spectro-radiometer) instrument (bands 27 to 36) onboard the AQUA polar satellite. We applied the fast Fourier transform (FFT) to these data as in Cavazzani et al. (\cite{cava17}).
A sharp periodicity of $178\pm7$ days is found, being June and December the most perturbed months.
We looked for shorter periodicities of the first harmonic, and the FFT analysis returned a periodicity of $14.0\pm0.5$ days of weaker atmospheric perturbations. 
This perturbation is most probably related to the 14-day luni-solar cycle (full moon and new moon). 
We then analysed the same Paranal database for pressure trends,  to search for physical explanations. The data are provided by the GLDAS Model (GLDAS datasets are available from the NASA Goddard Earth Sciences Data and Information Services Center (GES DISC). They give daily values for $1^{\circ}\times 1^{\circ}$ areas, like those of cloud cover data.
To this aim, we added the ocean tides at Chanaral De Las Animas (see Fig. \ref{i00}) database, and performed a triple correlation: atmospheric perturbations - lunar cycles - oceanic tides.
Fig. \ref{i00} shows the geographic area of interest and lists the characteristics of the various sites.
The layout of the paper is as follows.
In Section \ref{b} we briefly describe the Fourier Transform (FT) and in particular the Fast Fourier Transform (FFT) used in this analysis.
Section \ref{c} is devoted to the  analysis of  the correlation between atmospheric perturbations, oceanic tides and lunar cycles.
Section \ref{e}, finally, discusses the results and draws some conclusions.

\begin{figure}
  \centering
  \includegraphics[width=8cm]{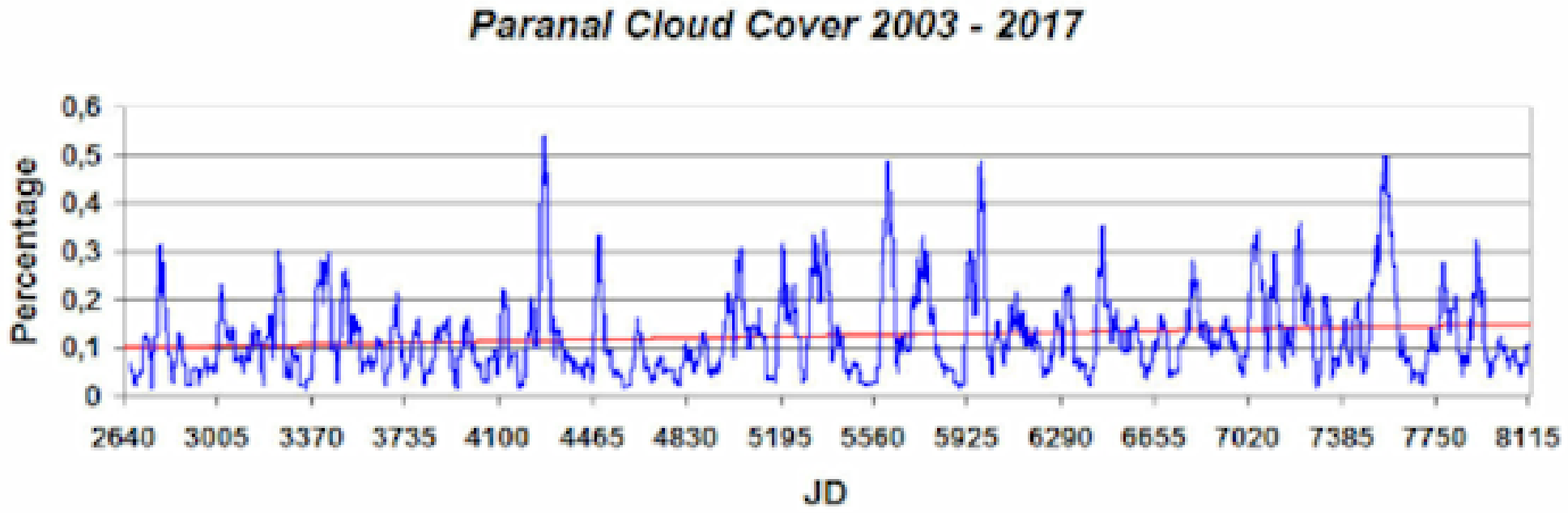}
  \includegraphics[width=8cm]{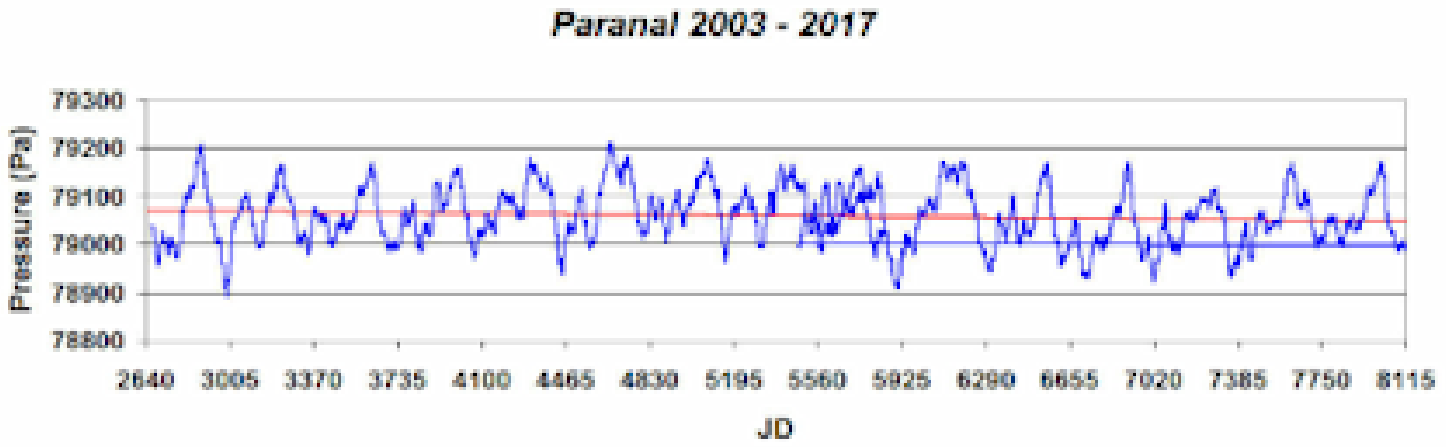}
  \caption{Top panel shows the cloud coverage trend at Paranal, bottom panel shows the pressure trend at Paranal during the same period. Time is in Julian days (JD): 2450000 + 2640 (x-axis).}
             \label{i01}
\end{figure}

\begin{figure}
  \centering
  \includegraphics[width=8cm]{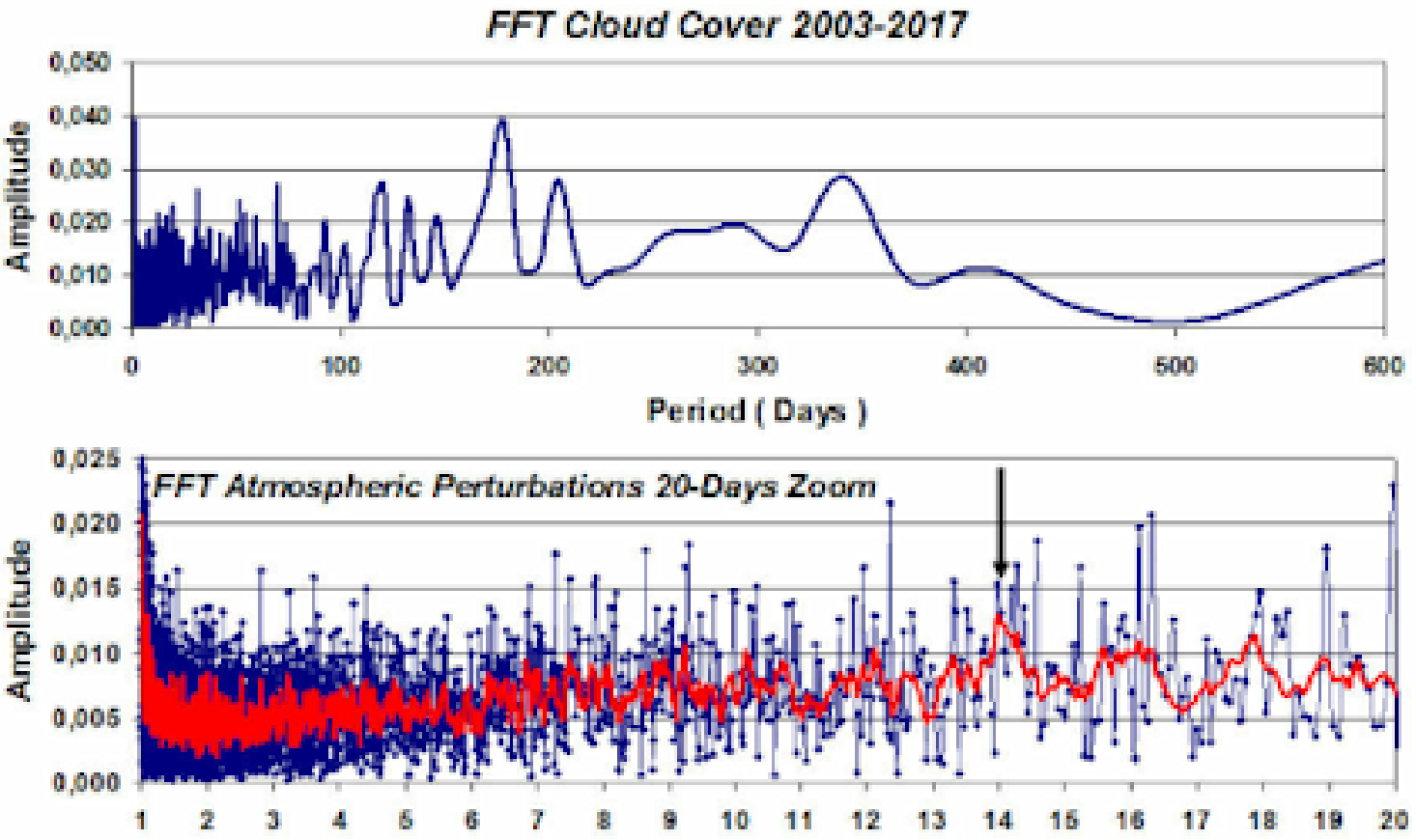}
  \includegraphics[width=8cm]{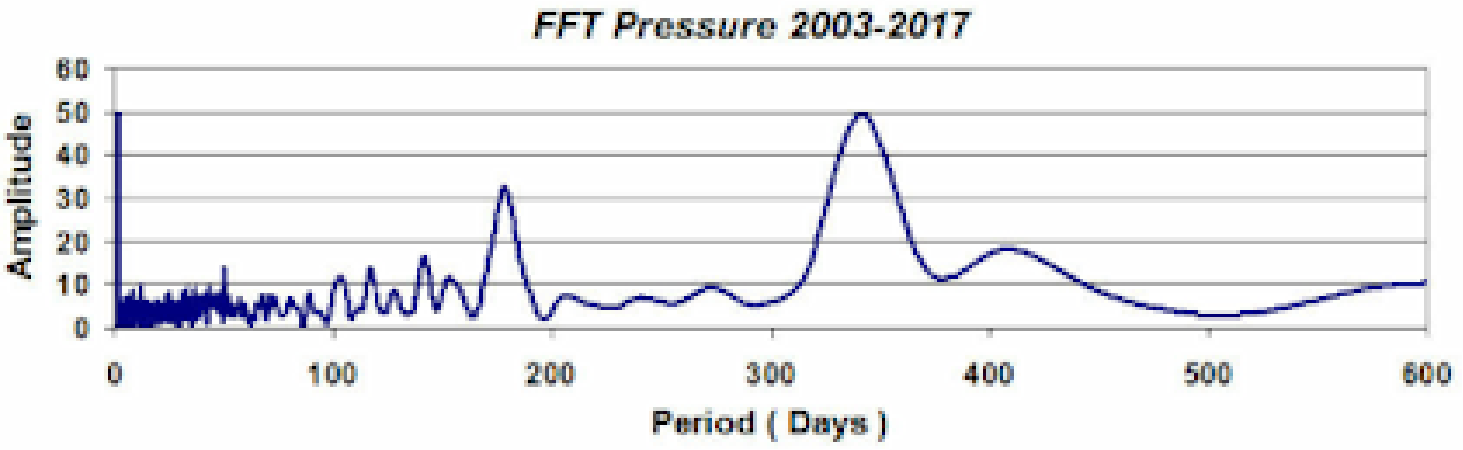}  
  \caption{(Top panel) FFT of the atmospheric perturbations (2003-2017 night MODIS data): main annual and semiannual periods.
(Central panel) FFT high-frequency zoom (blue) with its 7-point moving average curve (red): the highest 14-days peak is highlighted.
(Bottom panel) FFT of the pressure record at Paranal (2003-2017): main annual and semiannual periods.}
             \label{FFT}
\end{figure}

\section{Data analysis and error budget}
\label{b}

Following Cavazzani et al. (\cite{cava17}) we have applied the Fast Fourier Transform (FFT) to the cloud cover data (top panel) and the pressure data (bottom panel) at Paranal (2003-2017) presented in Fig. \ref{i01} as 28-day monthly averages from MODIS.
Fig. \ref{FFT} shows the respective FFT trend after the conversion of the main frequencies in main periods (Cooley-Tukey, \cite{ct}). 
We can see that we have a strong perturbation every 178 days corresponding to about half year (top panel).
The error analysis on the period of this peak is provided in three independent ways allowing a triple check of the validity of the results.
The first analysis is based on the error propagation and provides us with the maximum error due to data sampling.
Through the propagation formula described in Cavazzani et al. (\cite{cava17}) we get a  maximum error of 4$\%$,
which corresponds to an uncertainty of $\pm 7$ days.
The second error estimate is based on the frequency resolution of the Fourier analysis defined as $\nabla f=1/T=f_{P}/N$, where T is the period, $f_{P}$ is the frequency of the peak and N is the data population.
A frequency $f_{P}$ associated with a spectral peak has an uncertainty
of $\pm\frac{1}{2}\nabla f$ and the correspondent period P is:

\begin{equation}
	P=\frac{1}{f\mp\frac{1}{2}\nabla f}=\frac{1}{f}\pm\frac{\nabla f}{2f^{2}}=\frac{1}{f}\pm\frac{P^{2}}{2T}
	\label{for1}
\end{equation}

\begin{figure}
  \centering
  \includegraphics[width=8cm]{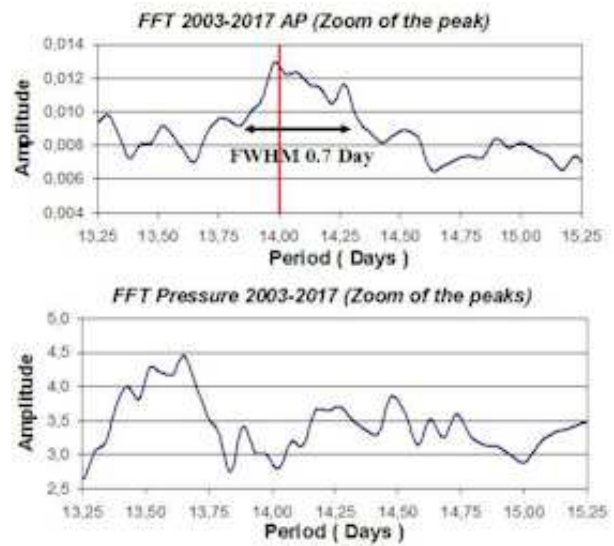}
  \caption{Comparison between the cloud cover periodicity and the periodicity of the pressure variation at Paranal. Top panel shows the zoom of the atmospheric perturbations peak with its associated FWHM. Bottom panel shows the zoom of the pressure peaks.}
             \label{compa}
\end{figure}

\begin{figure}
  \centering
  \includegraphics[width=8cm]{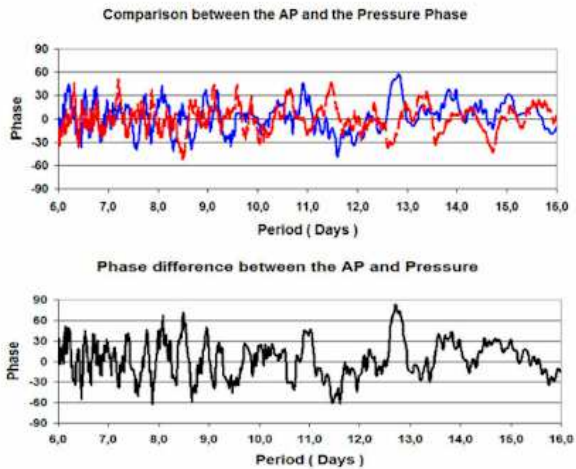}
  \caption{Comparison between the phase of atmospheric perturbations (continuous blue) and the phase of pressure oscillations (dashed red), the two oscillations are in phase on the 14th day.}
             \label{com}
\end{figure}

From this formula we derive an uncertainty of $\pm 4$ days on the semiannual period.
This result is consistent with the full width at half maximum (FWHM) of the analyzed peak.
The zoom in the top panel shows a FWHM of 10 days (hence, $\pm 5$ days). Thus, the uncertainty on the frequency resolution ($\pm 4$ days) and the FWHM ($\pm 5$ days) have a difference of one day only.
We believe that the most reliable error to be associated with the peak of 178 days is the maximum error due to the error propagation ($\pm 7$), which makes the peak compatible with a semiannual cycle.
The FFT spectra shown in Fig. \ref{FFT} also suggest the presence of minor oscillations with periods of about 206 and 412 days, which are also observed among the long-range oceanic tides (Avsyuk and Maslov, \cite{Avsyuk}). 
Central panel of Fig. \ref{FFT} shows the 20-days FFT high-frequency zoom (blue) with its 7-point moving average curve (red). This signal is correlated with the atmospheric perturbations (AP) at Paranal (Cavazzani et al., \cite{cava17}).
Fig. \ref{compa} shows FFT 7-points moving average spectra for the periods between 13.25 and 15.25 days (top panel). The main period of a perturbation at Paranal is about 14 days. 
Thus, there is a higher probability to have a covered night every two weeks: a fact that is validated by consulting the Paranal ground data on environmental monitoring\footnote{http://archive.eso.org/cms/eso-data/ambient-conditions.html}.
We did the same analysis with the pressure data at Paranal during the same period, as provided by the GLDAS Model.
Bottom panel of Fig. \ref{FFT} shows main semiannual and annual main period peaks in full agreement with the cloud cover results. 
The bottom panel of Fig. \ref{compa} shows a zoom of the FFT during the periods between 13.25 and 15.25 days: we have two periodicities. This perturbation is most probably related to the first harmonic of the tropical and synodical lunar months generating the lunisolar fortnightly $(M_{f})$ tide (period, 13.7 days) and the lunisolar synodic fortnightly $(M_{sf})$ tide (period, 14.6-15.0 days).
In order to strengthen our findings, we calculated the error associated with the 14-day peak in three different ways:
a maximum error of 4.0$\%$ which corresponds to 13 hours and 26 minutes through the error propagation due to satellite data; an error of 2.4$\%$ by applying the equation \ref{for1} which corresponds to 8 hours and 10 minutes and the 0.7-day FWHM of the 14-day peak corresponds to an error of 2.5$\%$ (8 hours and 24 minutes).
In summary we have obtained the following results: $14\pm0.56$ days with the error propagation, $14\pm0.34$ days with the frequency resolution of the Fourier analysis, and $14\pm0.35$ days with the FWHM of the 14-day peak (see Fig. \ref{compa}). This confirms the validity of the analysis.
The error on the amplitude of the peaks can be calculated through the formula $\Delta=\sigma^{2}/{T}$, where $\sigma$ is the statistical error of the satellite data and T is the analyzed period. Even if we consider a maximum error on the data of 10 percent and a period of 15 years (about 365 x 15 days) we obtain an amplitude error of about $\pm1.8\cdot10^{-4}$ which is negligible compared to those previously calculated.
Finally, we compared the phases of the two oscillations in Fig. \ref{com}, which  shows the phase of atmospheric perturbations (continuous blue) and the phase of pressure oscillations (dasched red), and again we note that the two oscillations are in phase on the 14th day.

\section{Correlation between atmospheric perturbations, oceanic tides and lunar cycles}
\label{c}

\begin{figure}
  \centering
  \includegraphics[width=8cm]{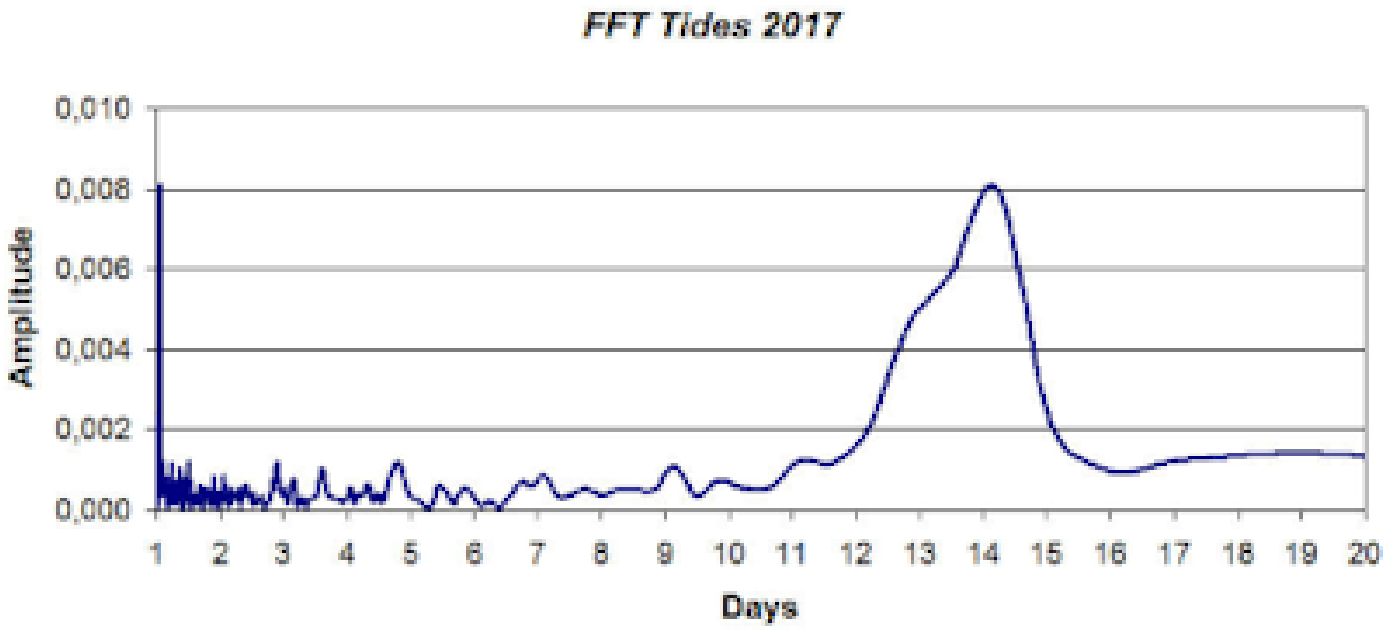}
  \includegraphics[width=8cm]{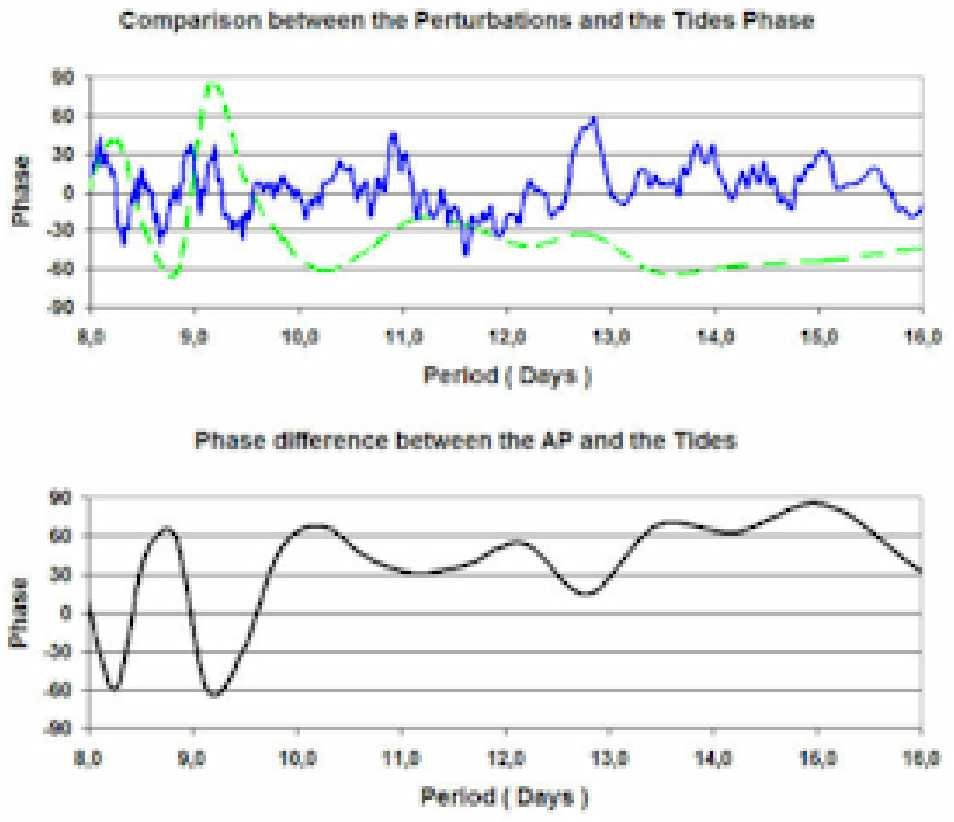}
  \caption{Top panel shows the FFT of the tides in Chanaral De Las Animas in 2017, central panel shows the comparison between the phase of atmospheric perturbations (continuous blue line) and the phase of the tides (dashed green line) and bottom panel shows the phase difference.}
             \label{com01}
\end{figure}

\begin{figure}
  \centering
  \includegraphics[width=8cm]{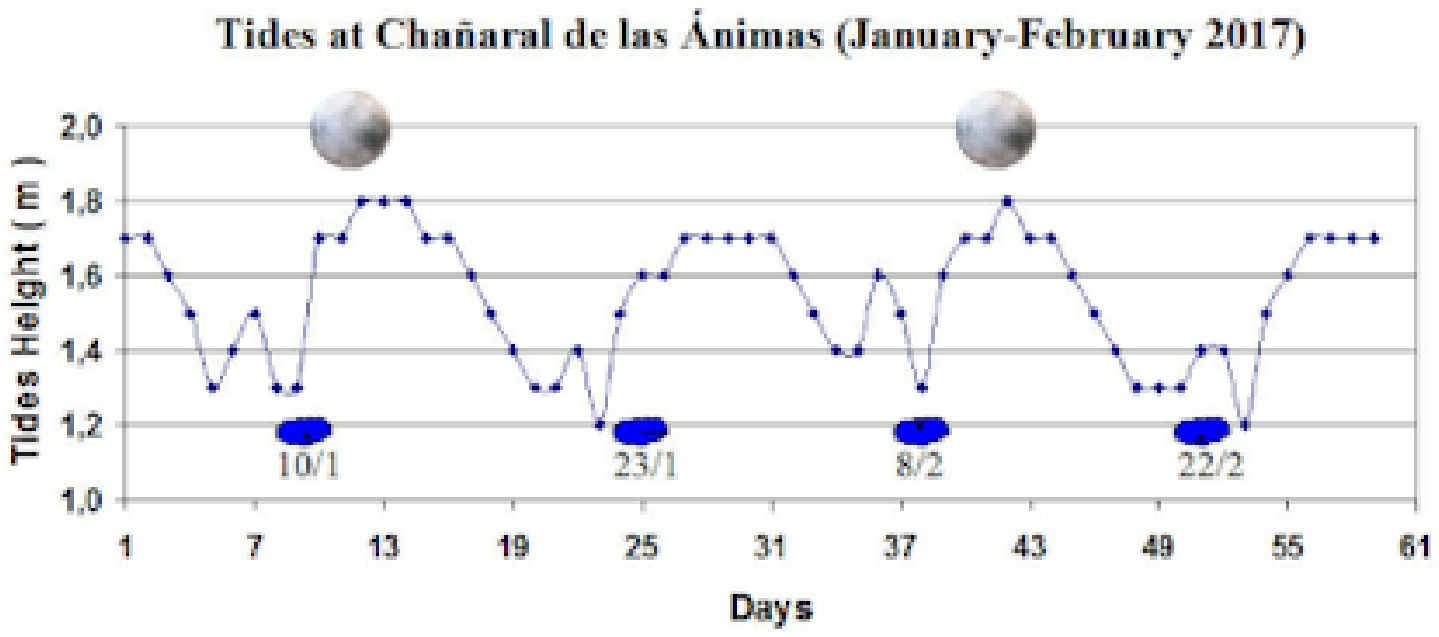}
  \includegraphics[width=8cm]{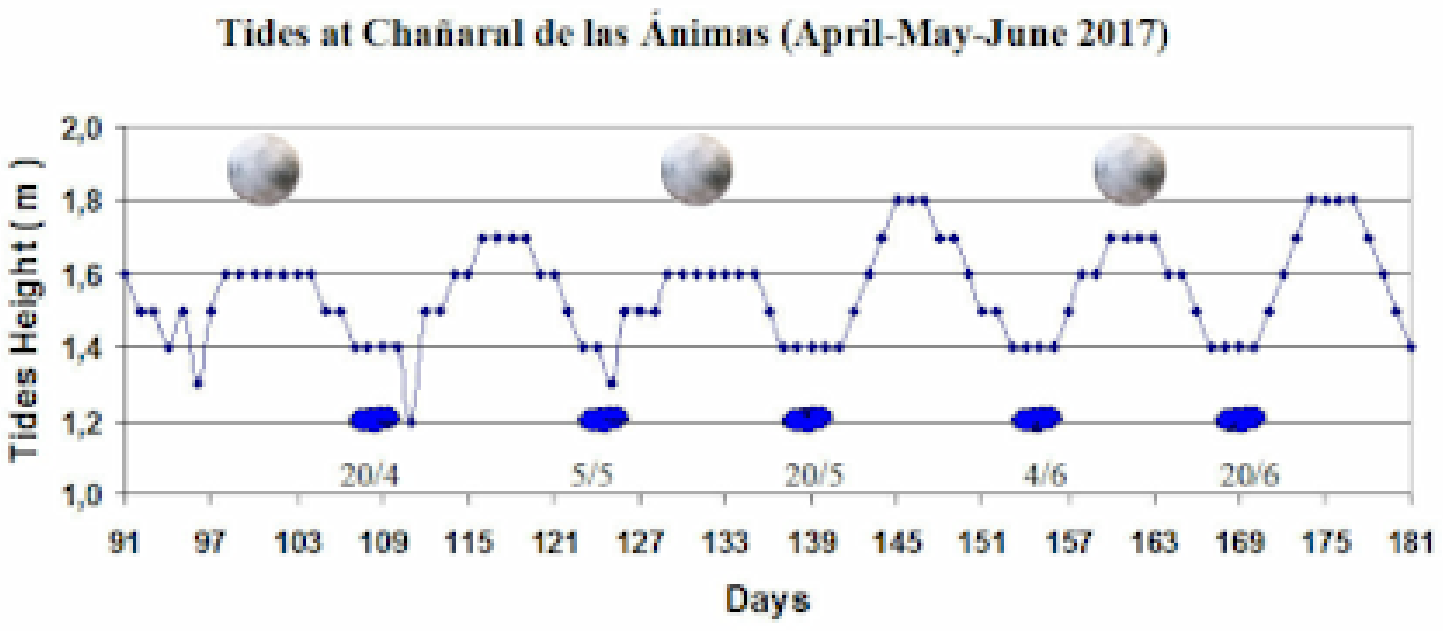}
  \caption{Tide maximum varies between $1.2 m$ and $1.8 m$ at Chanaral de las Animas, the low tide anticipates the full moon or the new moon of about three days. The figure shows an example of three types of atmospheric perturbations (represented schematically by clouds): an intense one with clouds and high seeing on 24 January, the second one is characterized by clouds, a worsening of the seeing and a precipitable water vapor (PWW) increase on 8 February and a third intense one with clouds and high seeing on 22 February, 2017.}
             \label{i03}
\end{figure}

In this Section we correlate the meteorological results with the ocean tides at Chanaral de las Animas (see Fig. \ref{i00}) and the underlying lunar cycles.
We extract from the 2017 tide database (Tides Planner app\footnote{https://www.imray.com/tides-planner-app/})
the daily maximum value of the tide.
The results of a FFT run on the tide data are shown in Fig. \ref{com01} (top panel). Very interestingly, one can clearly spot both the 14-day periodicity correlated with the lunar cycle and the 14-day oscillation of the atmospheric perturbations. The phases of the two fluctuations are calculated and compared in the central panel of Fig. \ref{com01}.
We noted that at the period of 14 days the tide phase is $-60$ degrees compared against the atmospheric perturbation phase, which translates into 2 days and 8 hours delay.
To cast more light on this evidence, we show in
Fig. \ref{i03} the trend of these data for January and February 2017. Atmospheric conditions data are taken from Paranal web page.
Over this period the tide maximum varies between $1.2$ m and $1.8$ m. The low tide anticipates the full moon or the new moon of about three days. 
A tight correlation seems to exists between the low tide (and hence moon) cycle and the atmospheric perturbations (indicated by blue clouds). 
Fig. \ref{i03} shows four different bad condition occurrences: on Jan 23 intense clouds and bad seeing, on Feb 8 clouds, seeing worsening and high precipitable water vapor (PWW),  and on Jan 10 and Feb 22 thick clouds and bad seeing.
We can see that the first two perturbations are separated by about 15 days, while the third is separated exactly by 14 days from the second. Besides, all perturbations occur at low 
tide.
Bottom panel shows the same trend for 2017 April, May and June. We can see 5 perturbations more in relation to the tides and the lunar cycle.
When extended to the whole years, this analysis yields a mean correlation as large as $96\%$.
Table \ref{cor} summarises the results of the correlation.
Column 2 shows the correlation between the atmospheric perturbations (AP) and the lunar cycles (LC): Fourier analysis shows that the AP anticipates the full moon or the new moon by 2 days and 8 hours. We considered the AP of January and analyzed the LC. If the perturbations occur with a temporal shift respect to the forecast the correlation decreases, i.e. we have about 2 perturbations every month. If these occur exactly 2 days and 8 hours before the full or new moon the correlation is 100 percent. A 3 percent error on a 31-day month corresponds to about 22 hours: this means that one perturbation can be anticipated or delayed by about one day, or both anticipate or delay by about 10 hours.
The same procedure was applied between the AP and the oceanic tides (OT) and between LC and OT columns 3 and 4 respectively.
We emphasize that our model is not aimed at an hourly precision: the model allows to forecast observation periods with a high probability of clear and stable atmospheric conditions with a long time horizon based on the moon phases.
At Chanaral de las Animas the minimum tide occurs between 5 pm and 7 pm (local time) during the days with atmospheric perturbations. 
This would take away superficial water from the ribs after the hot daily hours, consequently the coasts get cooler because the cold water rises from the depth and the temperature decreases would favour cloud formation or worsening of seeing.
Fried (1965) showed that seeing is related to the integral of $C^{2}_{n}$ (refractive index structure parameter). In turn, $C^{2}_{n}$ is linked to $C^{2}_{T}$ (temperature structure parameter) and therefore to the temperature variations.

\begin{table}
 \centering
 \begin{minipage}{80mm}
  \caption{Correlation between atmospheric perturbations (AP) at Paranal, lunar cycles (LC) and ocean tides (OT) at Chanaral (2017). The last column is the 3-month moving average (3-MMA).}
   \label{cor}
  \begin{tabular}{@{}lccccccccccc@{}}
  \hline

Paranal  &\multicolumn{4}{c}{AP-LC-OT correlations}		\\ 
  \hline
Month	 & AP-LC &	AP-OT & LC-OT & Mean & 3-MMA 	  \\
\hline
January	& 97&	97	& 100 & 98 & 95\\
February &	97	& 97	& 97	& 97 & 97 \\
March    	& 94	& 94	& 100	& 96 &97\\
April	& 93	& 97 & 100 & 97 & 96 \\
May&	94 &	97  &	97 &	96 & 95\\
June& 90	& 90 & 97 & 92 & 95 \\
July&	97 &	97	& 100 & 98 & 96\\
August& 97	& 100 & 100	& 99 & 98 \\
September& 93&	97	& 97	& 96 & 98\\
October	& 100&	100&	100&	100 & 97 \\
November	& 93 & 97 &	97 & 96	& 96 \\
December & 90 & 90 & 93 &	91 	& 95 \\
\hline
Mean & & & & 96 \\
\hline
\end{tabular}
\end{minipage}
\end{table}

\begin{figure}
  \centering
  \includegraphics[width=8cm]{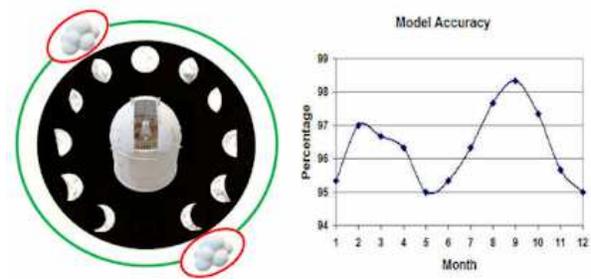}
  \caption{Left panel shows the lunar clock for the atmospheric perturbations forecast: at the top we see the perturbation during the Waxing Gibbons (WG) before the full moon while below we see the perturbation during the Waning Crescent (WC) before the new moon. The period between the two perturbations is stable with seeing smaller than 1 arcsec. Right panel shows the 3-month moving average for the accuracy of the model.}
             \label{lunar}
\end{figure}

\section{Discussion and conclusion}
\label{e}

In this paper we identified a weather perturbation at Cerro Paranal on a $\sim$14 days cycle which seems to be strongly correlated with the $M_{f}$ and $M_{sf}$ solar-lunar tidal cycles. 
In detail, we found that the cloud cover degree at Cerro Paranal increases during the low ocean tidal cycle on the Chilean coast at about 25° S latitude (see Fig~1). 
This happens close to the Waxing Gibbons (WB) and the Waning Crescent (WC) lunar phases that occur 3-4 days before the new or full moon, respectively (see Fig. \ref{lunar}). These perturbations normally last about one or two nights when the seeing index is larger than 1 arcsec. On the contrary, between the two perturbations periods the sky is normally clear with a seeing index smaller than 1 arcsec. This correlation varies with the season (see Table 1) and is maximum in October.  In general, the model accuracy is higher from February to April and from August to October (see Fig. \ref{lunar}, right panel). This result is very useful for the long-term observational program scheduling.
The solar-lunar tidal effects on the oceanic currents, as well as on meteorology (Hagan et al., 2003),  have been extensively discussed in the literature.  
Currie (1983) identified a periodic (18.6 years) nodal-induced drought in the Patagonian Andes. 
This 18.6 year cycle modulates the monthly and semi-monthly tidal cycles. 
Moreover, a  27.3 and 13.6 day cycles (produced by to solar-lunar tidal field) modulates the atmospheric circulation producing changes in zonal wind velocity (Li, 2005). 
A tidal forcing of the polar jet stream with periods of about 14 and 28 days tidal cycles has been recently observed (Best and Madrigali, 2016). Finally, in Shinsuke et al. (2015) a relationship is established between the 14-days cycle, the ocean temperatures, and the wind speed in Japan.
A possible physical explanation could be that on the Chilean coasts the Humboldt current combined with the tidal cycle activates a quasi 14-days recurrence of ocean water transport. At Chanaral de las Animas the minimum of low tide occurs in the late afternoon during the lunar phases of WG and WC. This dynamic favors the formation of thermal inhomogeneities along the Chilean coast and therefore a weather pertubation on the  Andes. This hypothesys could be tested in a following work.

\subsection{ACKNOWLEDGMENTS}

This activity is supported by the INAF (Istituto Nazionale di Astrofisica) funds allocated to the Premiale ADONI MIUR.
MODIS data were provided by the \textit{Giovanni - Interactive Visualization and Analysis} website.
We also refer to the 3D atmospheric reconstruction project at Prato Piazza (Italy).
Finally we thank the support of the University of Padua for this search (Research grant, type A, Rep. 138, Prot. 3022, 26/10/2018).

\label{lastpage}


\begin{thebibliography}{}










\bibitem[\protect\citeauthoryear{2011}{}]{Avsyuk}
Avsyuk, Y.N., Maslov, L.A., 2011, Earth Moon Planets, 108, 77


\bibitem[\protect\citeauthoryear{2016}{}]{best}
Best, C.H. and Madrigali, R., 2016, Atmos. Chem. Phys. Discuss., 15, 22701

\bibitem[\protect\citeauthoryear{2015}{}]{cava15}
Cavazzani, S., Ortolani, S., Zitelli, V., 2015, MNRAS, 452, 2185


\bibitem[\protect\citeauthoryear{2017}{}]{cava17}
Cavazzani, S., Ortolani, S., Zitelli, V., 2017, MNRAS, 471, 2616


\bibitem[\protect\citeauthoryear{1965}{}]{ct}
Cooley, J.W., Tukey J.W., 1965, Mathematics of Computation, 19, 297


\bibitem[\protect\citeauthoryear{1982}{}]{crawford}
Crawford, W.R., 1982, International Hydrographic Review, 59, 131


\bibitem[\protect\citeauthoryear{1983}{}]{currie}
Currie, R.G., 1983, Geophysical Research Letters, 10, 1089

\bibitem[\protect\citeauthoryear{1965}{}]{fried}
Fried D.L., 1965, J. Opt. Soc. Am., 55, 1427

\bibitem[\protect\citeauthoryear{2013}{}]{giorda}
Giordano C. et al., 2013, MNRAS, 430, 3102


\bibitem[\protect\citeauthoryear{2003}{}]{hagan}
Hagan, M.E., Forbes, J.M. and Richmond, A., 2003, Encyclopedia of Atmospheric Sciences, 1, 159.


\bibitem[\protect\citeauthoryear{2004}{}]{holton}
Holton, J.R., 2004, An Introduction to Dynamic Meteorology, Fourth Edition, Vol. 88 in the International Geophysics series


\bibitem[\protect\citeauthoryear{1998}{}]{Lakshmi}
Lakshmi, H., Kantha, J., Scott S., Shailen, D.D., 1998, Journal of Geophysical Reserrch, 103, 12639

\bibitem[\protect\citeauthoryear{2005}{}]{li}
Li, G., 2005, Adv. Atmos. Sci., 22, 359


\bibitem[\protect\citeauthoryear{2002}{}]{mascia}
Masciadri, E., Avila, R., Sanchez, L. J., 2002, Astronomy and Astrophysics, 382, 378


\bibitem[\protect\citeauthoryear{2018}{}]{Osborn}
Osborn, J., Sarazin, M., 2018, MNRAS, 480, 1278


\bibitem[\protect\citeauthoryear{2005}{}]{sara2005}
Sarazin, M., 2005, TMT Site Testing Workshop, Vancouver. 

\bibitem[\protect\citeauthoryear{2015}{}]{Shinsuke}
Shinsuke, I., Atsuhiko, I., Yasuyuki, M., 2015, Nature, Scientific Reports, 5, 10167


\end{thebibliography}
\end{document}